# Magnetic properties of the RbNd(WO$_4$)$_2$ single crystal

M.T. Borowiec[1*], E. Zubov[2], T. Zayarnyuk[1], M. Barański[1].

[1]Institute of Physics, Polish Academy of Sciences, al. Lotników 32/46, 02-668 Warsaw, Poland.

[2]A. A. Galkin Donetsk Physic-Technical Institute, Donetsk, Ukraine.

**Abstract**

The magnetic investigations as a function of temperature and magnetic field for the rubidium neodymium double tungstate RbNd(WO$_4$)$_2$ single crystal have been performed. The magnetization was measured in the temperature range from 4.2 to 100 K and for the magnetic field up to 1.5 T. The crystal field and exchange parameters were found.

[*] Corresponding author.
E-mail: borow@ifpan.edu.pl

## 1. Introduction

The rubidium neodymium tungstate RbNd(WO$_4$)$_2$ (RbNdW) is the representative of the family of alkaline (A) and rare earth (Re) double tungstates ARe(WO$_4$)$_2$ (AReW). RbNdW belongs to the monoclinic system with space group *C*2/*c*; which is isostructural with α-KY(WO$_4$)$_2$ [1]. This crystal is usually grown by high temperature solution method in order to obtain the low temperature monoclinic phase. At present, the AReW tungstates having the low-symmetric (e.g., monoclinic) crystalline structure and the atom arrangements in forms of chains or layers are intensively studied. Structural, optical and magnetic investigations of the KDyW, KHoW, KErW and RbDyW compounds were performed earlier [2]. Many of them show complicated structural phase transitions (SPT), caused by the cooperative Jahn–Teller effect (CJTE), and magnetic phase transitions.

The rubidium neodymium double tungstate, was also studied, especially its structural and spectroscopic properties [1].

In this paper, we show the new results of magnetization measurements for the rubidium neodymium double tungstate RbNd(WO$_4$)$_2$ (RbNdW) single crystal.

## 2. Magnetic properties of RbNd(WO$_4$)$_2$

The temperature, magnetic field and angular magnetization dependences of the RbNd(WO$_4$)$_2$ single crystal were investigated using the vibrating sample magnetometer (PAR Model 450) in a temperature range from 4.2 to 100 K for magnetic field up to 1.5 T. The field was applied both in the *ac* plane and along the *b*-axis.

The electron configuration of Nd$^{3+}$ is 4f$^3$. In the crystal field of monoclinic symmetry the ground multiplet $^4I_{9/2}$ splits into five Kramers doublets.

An angular dependence of magnetization has allowed to determine the magnetic *x* and *z*-axes, which correspond to the directions of minimal and maximal values of magnetization in the *ac* crystallographic plane, respectively (fig. 1). The angle between *c*- and *z*-axes in clockwise direction is equal to 86° and the angle between *a*- and *x*-axes is equal to 46°. The third main *y* magnetic axis is parallel to the second-order axis C$_2$ and coincides with the crystallographic *b*-axis perpendicular to the *ac* plane.

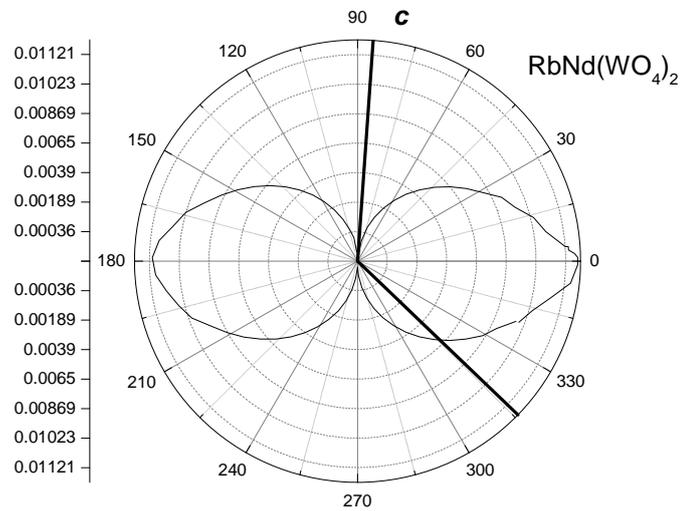

Fig. 1. The angular dependence of magnetization for RbNd(WO$_4$)$_2$ single crystal.

The magnetic field dependences of magnetization M$_i$ both along all magnetic axes and along the *x, y* and *z* crystal directions do not display the magnetization saturation in temperature interval from 4.2 to 100 K and in magnetic field up to 1.45 T (fig. 2), that it is characteristic for paramagnet.

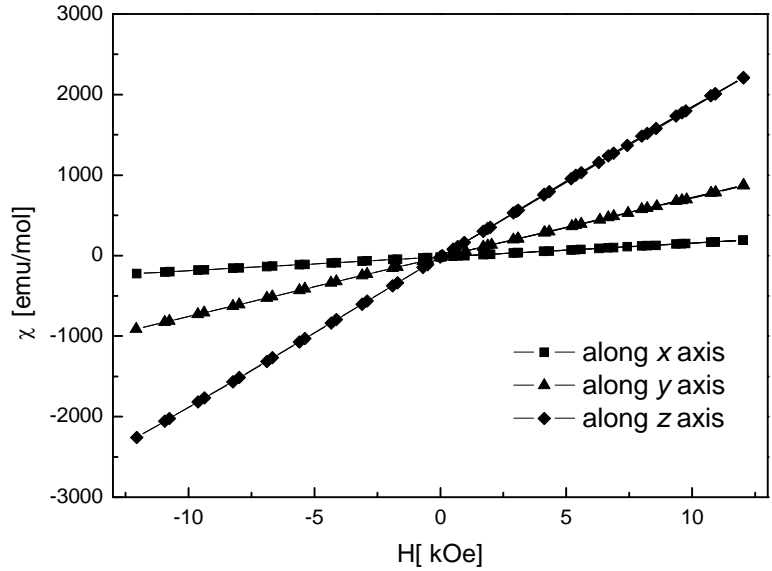

Fig. 2. The magnetic field dependences of magnetization along three magnetic axis at T=6 K.

For a weak magnetic field $H$ we may use linear relation for susceptibility $\chi_i(T) = M_i/H$. The experimental temperature dependences $\chi_i(T)$ were analyzed using the Curie-Weiss law:

$$\chi_i = \chi_{0i} + \frac{C_i}{T-\theta_i} \qquad (1)$$

General fitting was used for finding the parameters: $\chi_{0i}$ (temperature independent part of susceptibility), the Curie constant $C_i = N_A\,(\mu_B g_i)^2 J(J+1)/3k_B$ and $\theta_i$ (paramagnetic temperature) (see table 1).

Table 1. The fittig parameters for the Curie-Weiss law.

|  | $\chi_{0i}$ | $C_i$ | $g_i$ | $\theta_i$ |
|---|---|---|---|---|
| H‖a  H=0.125 T | $2.7\cdot10^{-3}$ emu/mol | 1.161 emu·K/mol | 3.52 | -105.13 K |
| H‖b  H=0.1 T | $1.06\cdot10^{-3}$ emu/mol | 5.55 emu·K/mol | 7.69 | -101.1 K |
| H‖c  H=0.1 T | $3.03\cdot10^{-3}$ emu/mol | 1.13 emu·K/mol | 3.59 | -0.654 K |

The results of fitting are also presented in Fig. 3(a,b,c). The paramagnetic temperature has a large value along the **x** and **y** directions.

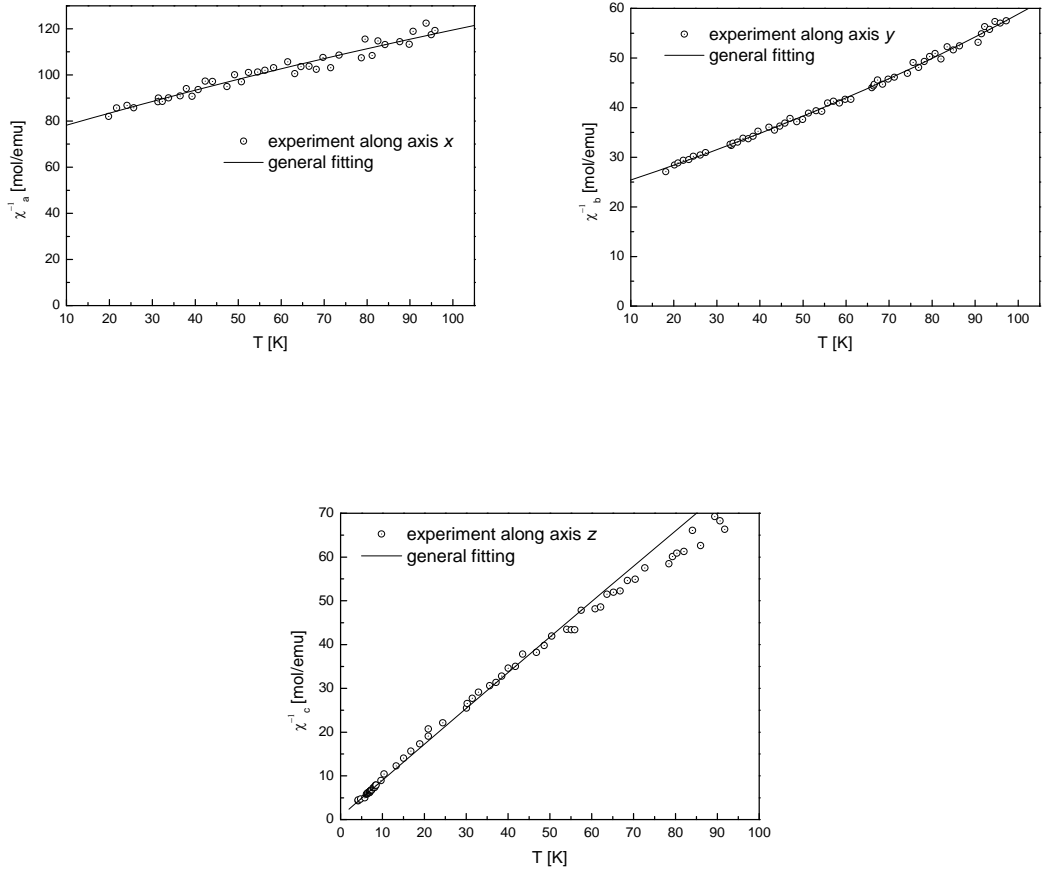

Fig. 3 abc. The temperature dependences of inverse susceptibility along three magnetic axes: points (experiment) and line (theory).

The obtained values of paramagnetic Curie temperature were used to find the crystal field parameters. For the $C_2$ symmetry of $Nd^{3+}$ site $\hat{H}_{cr}$ involves 15 non-zero crystal field parameters because for the z-axis parallel to the $C_2$ axis the crystal field parameters with the odd $q$ are equal to zero. Following to Ref. [3] we shall restrict consideration only to second order crystal field parameters.

Our system has the symmetry axis of the second order that excludes in crystal field Hamiltonian the Stevens operators with odd powers. Acording to that the Hamiltonian has the following form with the first nonvanishing terms:

$$\hat{H}_{cr.} = B_2^0 \left(3J_z^2 - J(J+1)\right) + B_2^2 \left(J_x^2 - J_y^2\right) \qquad (2)$$

We attempted to estimate crystal field parameters $B_2^0$ and $B_2^2$. At high temperatures, the relations between the crystal field parameters and crystal field contribution in paramagnetic temperatures $\theta_i^{cr}$ in the $i$ direction have a form

$$\theta_x^{cr.} = \frac{1}{2}\xi\left(B_2^0 - B_2^2\right)$$
$$\theta_y^{cr.} = \frac{1}{2}\xi\left(B_2^0 + B_2^2\right) \quad , \quad (3)$$
$$\theta_z^{cr.} = -\xi B_2^0$$

where $\xi = \frac{1}{5}(2J-1)(2J+3)$ [3]. For J = 9/2 $\xi$ = 96/5. In our case, the *x, y* and *z* axes coincide with the *c, a* and *b* axes, respectively. Then the paramagnetic temperature $\theta_i$ can be expressed as [3]

$$\theta_i = \theta_i^{cr.} + \theta_i^{exch.}, \quad (4)$$

where the exchange contribution in paramagnetic temperature is equal to $\theta^{exch.} = \frac{2}{3}J(J+1)J(0) = \frac{33}{2}J(0)$. The J(0)=z*J, where z -number of the nearest neighbours of the rare-earth ion. J is the parameter of the pair exchange interaction.

Since the contribution of crystal field in sum of paramagnetic temperatures along the 3rd direction is equal to zero, only the 3rd contribution from exchange remain.

From relation $\theta_x + \theta_y + \theta_z = \frac{99}{2}J(0)$ we obtain J(0) = -4.18 K, and the crystal field contribution in paramagnetic temperatures is equal to $\theta_x^{cr.} = 68.3K$, $\theta_y^{cr.} = -36.16K$ and $\theta_z^{cr.} = -32.16K$.

Then we obtain the following parameters of crystal field Hamiltonian $B_2^0 = 1.68K$, $B_2^2 = -5.44K$. It gives five Kramers doublets with energies of 0, 85.4, 149.4, 191.9 and 213.9 K.

In summary, we have found that the magnetization shows a strong anisotropy in temperature range studied. In the temperature range up to 100 K, the experimental curves of susceptibility follow to the Curie-Weiss law. By fitting the calculated susceptibility to the experimental data, the crystal field and exchange parameters and g-factors along main crystal directions were calculated. The energies of $Nd^{3+}$ spectra in a low symmetric crystal field were estimated.

**Acknowledgements**

This work was supported by EU project DT-CRYS, NMP3-CT-2003-505580, of the Polish State Committee for Scientific Research (KBN) (decision of project No. 72/E-67/SPB/6. PR/DIE 430/2004-2006) , and by the European Regional Development Fund through the Innovative Economy grants POIG 01.03.01-00-058/08 and POIG 01.01.02-00-108/09.

**Reference**

[1] M.T. Borowiec, A.D. Prokhorov, I.M. Krygin, V.P. Dyakonov, K. Wozniak, L. Dobrzycki, T. Zayarnyuk, M. Baranski, W. Domukhowski, H. Szymczak, Physica B **371**, 205 (2006)

M.T.Borowiec, A.A.Prokhorov, A.D.Prokhorov, V.P.Dyakonov, H.Szymczak J. Phys., Condens. Matter **15**, 5113 (2003)

G. Leniec, T. Skibinski, S.M. Kaczmarek, P. Iwanowski, M. Berkowski Cent.Eur.J.Phys. **10**, 500 (2012)

[2]. M.T. Borowiec, A. Watterich, T. Zayarnyuk, V.P. Dyakonov, A. Majchrowski, J. Zmija, M. Baranski, H. Szymczak, J. Appl. Spectrosc. **71**, 888 (2004)

M.T. Borowiec, I. Krynetski, V.P. Dyakonov, A. Nabiałek, T. Zayarnyuk, H. Szymczak, New J. Phys.**8**, 124 (2006)

M.T. Borowiec, V.P. Dyakonov, K.Wozniak, L. Dobrzycki, M. Berkowski, E.E. Zubov, E. Michalski, A. Szewczyk, M.U. Gutowska, T. Zayarnyuk, H. Szymczak, J. Phys., Condens. Matter **19**, 056206 (2007)

M.T. Borowiec, Proc. SPIE **4412**, 196 (2001)

M.T. Borowiec, A.D. Prokhorov, I.M. Krygin, V.P. Dyakonov, K. Wozniak, L. Dobrzycki, T. Zayarnyuk, M. Baranski, W. Domukhowski, H. Szymczak, Physica B **371**, 205 (2006)

[3] A.K.Zvezdin, V.M.Matveev, A.A.Mukhin, A.I.Popov, "Rare earth ions in magnetically ordered crystals", Moscow, Nauka, 1985, p.120.